\documentstyle[12pt]{article}
\begin{document}
\setlength{\unitlength}{1mm}
\begin{center}
{\Large\bf Supersymmetric Topological Quantum Field Theories
Of Differential Forms I. Gauge p-forms}
\end{center}
\begin{center}
{\large\bf S.N.Solodukhin}
\end{center}
\begin{center}
{\bf Department of Theoretical Physics, Physics Faculty of Moscow
University, Moscow 117234, Russia}
\end{center}
\vspace*{2cm}
\abstract
We consider the topological theory of Witten type for gauge differential
p-forms. It is shown that some topological invariants such as linking
numbers  appear under quantization of this theory. The non-abelian
generalization of the model is discussed.
\vspace*{2cm}
\section{Introduction}
\setcounter{equation}0

   Recently much attention has been payed to the study of topological
quantum field theories ~\cite{ew88},~\cite{gh89}, ~\cite{as78},
{}~\cite{pr91}, ~\cite{gh90}, ~\cite{sw91}. In these theories the action
functional either doesn't depend on metric on the space-time manifold
$M^{n}$ or metrical variation of action is BRST-commutator. Such a
theories are classified as correspondingly of Schwarz ~\cite{as78} and
Witten ~\cite{ew88} classes. One obtains the topological invariants under
consideration the quantum average of operators independent on space-time
metric and BRST-invariant in Witten case ~\cite{ew88}.

   The wide class of Schwarz type theories was considered in
{}~\cite{gh89} by means of differential forms on $M^{n}$. The simplest
action of this type has the form ~\cite{gh89},~\cite{as78}:
\begin{equation}
S = \int_{M^{n}}^{ } B \wedge dA
\label{eq:1.1}
\end{equation}
where $B$ and $A$ are correspondingly rank $p$ and $n-p-1$
differential forms. The action (1.1) is invariant under gauge
transformations:
\begin{eqnarray}
&&A \rightarrow A + d \alpha \nonumber\\
&&B \rightarrow B + d \beta
\label{eq:1.2}
\end{eqnarray}
where $\alpha$ and $\beta$ are differential forms of rank $(n-p-2)$
and $(p-1)$ correspondingly. The space of gauge inequivalent solutions
for the action (1.1) is direct sum of cohomologies $H^{(p)}(M) \oplus
H^{(n-p-1)}(M)$ ~\cite{gh89}.

   Let $U_{p}$ and $U_{n-p-1}$ are two nonintersecting submanifolds
of dimension $p$ and $(n-p-1)$, which represent an trivial elements in
$H_{p}(M)$ and $H_{(n-p-1)}(M)$ correspondingly. Then one obtains that
the quantum average values
\begin{equation}
< \int_{U_{p}}^{ } B \int_{U_{n-p-1}}^{ } A > = L(U_{p}, U_{n-p-1})
\label{eq:1.3}
\end{equation}
give us linking number $L(U_{p}, U_{n-p-1})$ of manifolds $U_{p}$
and $U_{n-p-1}$ ~\cite{gh90}, ~\cite{sw91}.

   On the other hand the following action is well known ~\cite{yn76}
\begin{equation}
S = \int_{M^{n}}^{ } {1 \over 2} \ast dA_{p} \wedge dA_{p}
\label{eq:1.4}
\end{equation}
(where we denote $A_{p}$ the rank p differential form),which is also invariant
under the (1.2)-type gauge transformations:
\begin{equation}
A_{p} \rightarrow A_{p} +d \alpha_{p-1}
\label{eq:1.5}
\end{equation}
However the action (1.4) depends on metric via the Hodge
dualization operator $\ast$.

   The other similar action is following
\begin{equation}
S = \int_{M^{n}}^{ } {1 \over 2} \ast \delta B_{p} \wedge \delta B_{p}
\label{eq:1.6}
\end{equation}
where $\delta$ is operator conjugated to $d$ with respect to natural
scalar product $\int_{ }^{ } \ast C_{p} \wedge B_{p}$. Action (1.6) is
invariant under  following gauge transformations:
\begin{equation}
B_{p} \rightarrow B_{p} + \delta \alpha_{p+1}
\label{eq:1.7}
\end{equation}
where $\alpha_{p+1}$ is arbitrary (p+1)-form.

   It is easy to see that the space of gauge inequivalent solutions
for the actions (1.4), (1.6) is cohomology space $H^{p}(M)$. The
actions (1.4), (1.6) are not topological since they depend on metric.
However one can attempt to get topological action of Witten type by
adding an corresponding fermion term to (1.4), (1.6). The action of
this type in two dimensions was constructed in ~\cite{ss92} to describe
the nonstandard string model. Here we consider such a supersymmetric
extension in any dimensions. As it was noted in ~\cite{yn76} the action
(1.6) can be transformed to the form (1.4) by substituting
$A_{n-p}=\ast B_{p}$. Therefore we will consider only the action (1.4).

\section{The Supersymmetric Action}
\setcounter{equation}0
   Let us consider the boson differential p-form $A_{p}$ and set of
fermion forms $\Psi_{p}$ and $\Psi_{p+1}$, valued in Grassman algebra
( two such Grassman forms $\psi_{p}$ and $\chi_{q}$ satisfy $\psi_{p} \chi_{q}
=
{(-1)}^{qp+1} \chi_{q} \phi_{p}$, which differs by a minus sign from the
relation satisfied by conventional bosonic forms: $A_{p} B_{q}={(-1)}^{pq}
B_{q}A_{p}$ ).

   The supersymmetric generalization of
action (1.4) is following
\begin{equation}
S^{(p)} = \int_{M^{n}}^{ } { 1 \over 2} \ast dA_{p} \wedge dA_{p} +
\ast \Psi_{p+1} \wedge d \Psi_{p}
\label{eq:2.1}
\end{equation}
(We write the fermion part of action (2.1) in this form in order to
the supersymmetry transformation (see eq.(2.3)) would not depend on
metric over $M^{n}$).

   It should be noted that action of this type was considered recently in
{}~\cite{cp91} from other point of view ( the action in ~\cite{cp91}
corresponds to our one in the particular case $n=4k+1$, $p=2k$ and under
constraint $\ast \Psi_{p+1}=\Psi_{p}$).

   The action (2.1) is invariant under gauge transformations:
\begin{eqnarray}
&&A_{p} \rightarrow A_{p} + d\alpha_{p-1}\nonumber\\
&&\Psi_{p} \rightarrow \Psi_{p} + d\eta_{p-1}\nonumber\\
&&\Psi_{p+1} \rightarrow \Psi_{p+1} + \delta \xi_{p+2}
\label{eq:2.2}
\end{eqnarray}
where $\eta_{p-1}$ and $\xi_{p+2}$ are Grassman differential forms
of rank (p-1) and (p+2) correspondingly.

   The action (2.1) is invariant also under transformation of
supersymmetry:
\begin{eqnarray}
&&\delta F = \epsilon \hat{Q} F \equiv \epsilon \{ Q, {\cal F} \}\nonumber\\
&&\hat{Q} A_{p} = \Psi_{p}\nonumber\\
&&\hat{Q} \Psi_{p+1} = -dA_{p}
\label{eq:2.3}
\end{eqnarray}
Generally the operator $\hat{Q}$ is not nilpotent since we have
\begin{eqnarray}
&&{\hat{Q}}^{2}  \Psi_{p+1} =  \hat{T} \Psi_{p+1}\nonumber\\
&&\hat{T} \Psi_{p+1} = -d \Psi_{p}
\label{eq:2.4}
\end{eqnarray}
Consequently operator $\hat{Q}$ is nilpotent only on space of
$\hat{T}$-invariant functionals. Notice that though the action (2.1)
is $\hat{T}$-invariant it hasn't the form of $\hat{Q}$-commutator. One
should introduce an additional boson (p+1)-form $A_{p+1}$ with
following transformation law under the supersymmetry:
\begin{equation}
\hat{Q} A_{p+1} = \Psi_{p+1}
\label{eq:2.5}
\end{equation}
Then the action $S^{(p)}$ has the form
\begin{eqnarray}
&&S^{(p)} = \hat{Q} X \nonumber\\
&&X= \int_{ }^{ } -{1 \over 2} \ast \Psi_{p+1} \wedge dA_{p} + { 1 \over 2}
\ast A_{p+1} \wedge d \Psi_{p}
\label{eq:2.6}
\end{eqnarray}
Acting by operator ${\hat{Q}}^{2}$ on $A_{p+1}$ one obtains
\begin{eqnarray}
&&{\hat{Q}}^{2} A_{p+1} = \hat{T} A_{p+1} \nonumber\\
&&\hat{T} A_{p+1} = - dA_{p}
\label{eq:2.7}
\end{eqnarray}
One can see that the functional X (2.6) is $\hat{T}$-invariant and it
is invariant also under gauge transformations of $A_{p+1}$:
\begin{equation}
A_{p+1} \rightarrow A_{p+1} + \delta \beta_{p+2}
\label{eq:2.8}
\end{equation}
It is essential in Witten theory ~\cite{ew88} that energy-momentum
tensor is $\hat{Q}$-commutator. We also have for variation of (2.1)
under the metric $g$:
\begin{equation}
\delta_{g} S^{(p)} = \hat{Q} ( \delta_{g} X)
\label{eq:2.9}
\end{equation}
Let us note that though the additional field $A_{p+1}$ doesn't
contribute explicitly into the action it is part of supermultyplet.
Hence one should take into account the field $A_{p+1}$ and gauge
invariance with respect to (2.8) in functional integral under
quantization.
   One gets from (2.1) the classical equations of motion:
\begin{equation}
\delta d A_{p} = 0, \\ \delta \Psi_{p+1} = 0, \\ d \Psi_{p} = 0
\label{eq:2.10}
\end{equation}
Hereafter we will consider $M^{n}$ be compact closed manifold. Then
modulo gauge transformations (2.2) the solutions of equations (2.10)
are represented by elements of cohomology spaces: $A_{p}\in H^{p}(M),
\Psi_{p}\in H^{p}(M), \Psi_{p+1} \in H^{p+1}(M)$.

\section{The Classical Hodge Theory}
\setcounter{equation}0
   According to Hodge decomposition the space $\Omega^{p}(M)$ on
manifold $M$ is decomposed on direct sum:
\begin{equation}
\Omega^{p}(M) = d \Omega^{p-1} \oplus \delta \Omega^{p+1}(M) \oplus
Ker \Delta_{p}
\label{eq:3.1}
\end{equation}
where $\Delta_{p}$ is Laplacian acting on the space of p-forms:
\begin{equation}
\Delta_{p} = \delta_{p+1} d_{p} + d_{p-1} \delta_{p}
\label{eq:3.2}
\end{equation}
The elements of the space $Ker \Delta_{p}$ are p-forms $\phi_{0}^{(p)} \in
H^{p}(M)$ satisfying the equations:
\begin{equation}
d_{p} \phi_{0}^{(p)} = \delta_{p} \phi_{0}^{(p)} = 0
\label{eq: 3.3}
\end{equation}
In the space $d \Omega^{p-1}(M)$ there is natural basis
{$\beta_{n}^{(p)}$} of eigenforms of the Laplacian $\Delta_{p}$:
\begin{equation}
d \delta \beta_{n}^{(p)} = [\tilde{\lambda}_{n}^{(p)}]^{2}
\beta^{(p)}_{n}
\label{eq:3.4}
\end{equation}
where eigenvalues $\{\tilde{\lambda}_{n}^{(p)} \}$ may depend on rank
p.

   On the other hand there is corresponding basis $\{ \alpha_{n}^{(p)} \}$
on the space $\delta \Omega^{p-1}(M)$:
\begin{equation}
\delta d \alpha_{n}^{(p)} = [\lambda_{n}^{(p)}]^{2} \alpha_{n}^{(p)}
\label{eq: 3.5}
\end{equation}
The basis (3.4), (3.5) corresponds to nonzero (positive) eigenvalues
of $\Delta_{p}$ (3.3).
   One can see that if p-form $\beta_{n}^{(p)}$ is solution of
eq.(3.4) then (p-1)-form $\alpha_{n}^{(p)} =
(\tilde{\lambda}_{n}^{(p)})^{-2} \delta \beta^{(p)}_{n}$ is solution
of eq.(3.5): $\delta d \alpha^{(p-1)}_{n} =
[\tilde{\lambda}_{n}^{(p)}]^{2} \alpha^{(p-1)}_{n}$. And vice versa,
if $\alpha_{n}^{(p)}$ is solution of (3.5) then (p+1)-form
$\beta_{n}^{(p+1)} = (\lambda_{n}^{(p+1)})^{-2} d\alpha_{n}^{(p)}$ is
solution of (3.4): $d \delta \beta_{n}^{(p+1)} = [\lambda_{n}^{(p)}]^{2}
\beta^{(p+1)}_{n}$.
Hence one gets the identity for eigenvalues:
\begin{equation}
\tilde{\lambda}_{n}^{(p)} = \lambda_{n}^{(p-1)}
\label{eq: 3.6}
\end{equation}
The Hodge operator $\ast$: $\Omega^{p}(M) \rightarrow \Omega^{(n-p)}(M)$
satisfies the identity $\ast \ast = (-1)^{p(n-p)}$. On the other hand the
adjoint of $d_{p-1}$ operator is $\delta_{p} = (-1)^{pn+n+1}
\ast d_{n-p} \ast$. Then one obtains that
\begin{equation}
\ast d\delta = \delta d \ast
\label{eq: 3.7}
\end{equation}
Consequently, if $\alpha_{n}^{(p)}$ is solution of (3.5) with
eigenvalue $\lambda_{n}^{(p)}$ then $(n-p)$-form $\ast \alpha_{n}^{(p)}$
satisfies eq.(3.4) with the same eigenvalue, i.e. $\beta_{m}^{(n-p)} =
\ast \alpha^{(p)}_{m}$ and $\tilde{\lambda}_{m}^{(n-p)} =
\lambda_{m}^{(p)}$. Together with (3.6) it gives us that
\begin{equation}
\lambda_{m}^{(n-p-1)} = \lambda_{m}^{(p)}
\label{eq: 3.8}
\end{equation}

   Let $\{ \alpha_{n}^{(p)} \}$ is orthonormal set of eigenforms (3.5)
then $\{ \beta_{n}^{(p)} = {1 \over {\lambda_{n}^{(p-1)}}} d
\alpha_{n}^{(p-1)} \}$ is othonormal set of eigenforms (3.4).

   We have the canonical decomposition for arbitrary p-form:
\begin{equation}
A_{p} = \sum_{i=1}^{\beta_{p}} b_{i}^{(p)} \phi_{i}^{(p)} + \sum_{n}^{
} a_{n}^{(p)} \alpha_{n}^{(p)} + \tilde{a}_{n}^{(p)} \beta_{n}^{(p)}
\label{eq: 3.9}
\end{equation}
where $\phi_{i}^{(p)}$, $i= 1, ... , \beta_{p}$ is basis of harmonic
p-forms, $\beta_{p}$ is p-th Betti number.
   One has similar decomposition for fermion p-form:
\begin{equation}
\Psi_{p} = \sum_{i_{p}}^{\beta_{p}} \eta_{i}^{(p)} \phi_{i}^{(p)} +
\sum_{n}^{ } \theta_{n}^{(p)} \alpha_{n}^{(p)} +
\tilde{\theta}_{n}^{(p)} \beta_{n}^{(p)}
\label{eq: 3.10}
\end{equation}
where the coefficients are anticommutating.
   Subsituting now (3.9), (3.10), into action (2.1) one gets:
\begin{equation}
S^{(p)} = \sum_{n}^{ } { 1 \over 2} [\lambda_{n}^{(p)}]^{2}
a_{n}^{(p)} a_{n}^{(p)} + \lambda_{n}^{(p)} \tilde{\theta}_{n}^{(p+1)}
\theta_{n}^{(p)}
\label{eq:3.11}
\end{equation}
It is easy rewrite the supersymmetry transformations (2.3), (2.5) in
terms of new variables:
\begin{eqnarray}
&&\hat{Q} a^{(p)}_{n} = \theta_{n}^{(p)}\nonumber\\
&&\hat{Q} \tilde{\theta}_{n}^{(p+1)} = - \lambda_{n}^{(p)}
a_{n}^{(p)}\nonumber\\
&&\hat{Q}\tilde{a}_{n}^{(p+1)} = \tilde{\theta}_{n}^{(p+1)}
\label{eq:3.12}
\end{eqnarray}
It is worth noting that the variation of action $S^{(p)}$ (2.1) under
metric is not $\hat{Q}$-commutator (without introduction of additional
field $A_{p+1}$). Nevertheless one gets for variation of $S^{(p)}$
(3.11) under variation of eigenvalues $\lambda_{n}^{(p)}$:
\begin{equation}
\delta_{\lambda} S^{(p)} = \sum_{n}^{ } \delta {\lambda_{n}^{(p)}}
\hat{Q}(- \tilde{\theta}_{n}^{(p+1)} a_{n}^{(p)} )
\label{eq:3.13}
\end{equation}

\section{Quanization And Topological Invariants}
\setcounter{equation}0
   We fix  the gauge
\begin{equation}
A^{L}_{p} = 0, \Psi_{p}^{L} =0, A^{T}_{p+1}=0, \Psi^{T}_{p+1}=0
\label{eq:4.1}
\end{equation}
where $B^{L}_{p}\in d\Omega^{p-1}(M), B^{T}_{p} \in \delta
\Omega^{p+1}(M)$.
   We get the following measure on the space of gauge inequivalent
forms $A_{p}$ (for more details see ~\cite{as78}, ~\cite{as84}):
\begin{equation}
{\cal D} A_{p} = {\cal D} A^{T}_{p} {\cal F}_{p} d\mu_{0}^{(p)},
\end{equation}
where
\begin{eqnarray}
&&{\cal D} A^{T}_{p} = \prod_{n} da_{n}^{(p)} det^{1 \over 2}
(<\alpha_{n}^{(p)},
\alpha^{(p)}_{m} >),\nonumber\\
&&< \phi^{(p)}, \psi^{(p)}> = \int_{M^{n}} \ast \phi^{(p)} \wedge \psi^{(p)}
\end{eqnarray}
is integrating measure on the space of the transverse p-forms $\delta \Omega^{
p+1} (M)$.We have the standard form  ~\cite{as78},~\cite {yn76},~\cite {as84}
for the
factor ${\cal F}_{p}$ in (4.2) due to gauge fixing:
\begin{eqnarray}
&&{\cal F}_{p} = E^{1 \over 2}_{p-1} E^{-{1 \over 2}}_{p-2} ...\nonumber\\
&&E_{p-i} = det'( \delta_{p-i+1} d_{p-i})
\end{eqnarray}
where $det'$ denotes the zeta-function regularization of determinant with zero
modes excluded.
In the integration measure over zero modes $d\mu_{0}^{(p)}$ one should take
into account both zero-modes
of p-form $A_{p}$ and zero-modes of operators $\delta_{p-i+1} d_{p-i}$ (4.4):
\begin{eqnarray}
&&d \mu_{0}^{(p)} = \prod_{i=1}^{\beta_{p}} db_{i}^{(p)} det^{1 \over 2}
(<\phi^{(p)}_{i}, \phi^{(p)}_{j}>)\nonumber\\
&&\prod_{k=1}^{p} \prod_{i=1}^{\beta_{p-k}} df_{i}^{(p-k)}
[\det(<\phi_{i}^{(p-k)}, \phi^{(p-k)}_{j}>)]^{(-1)^k \over 2}
\end{eqnarray}
where we introduce the ghost zero-modes $f_{i}^{(p-k)}$ which are fermionic for
odd k
and bosonic for even k.

   As it was noted above we have to take into account the integration over
additional field $A_{p+1}$
with gauge invariance under (2.8) in functional integral:
\begin{equation}
{\cal D}A_{p+1} = {\cal D} A^{L}_{p+1} \tilde{{\cal F}}_{p} d \tilde{
\mu}_{0}^{(p+1)}
\end{equation}
where
\begin{equation}
{\cal D} A^{L}_{p+1} = \prod_{n} d \tilde{a}^{(p+1)}_{n} det^{1 \over 2}
(<\beta^{(p+1)}_{n}, \beta^{(p+1)}_{m}>)
\end{equation}
is measure on the space of longitudinal forms $d \Omega^{p}(M)$.
   We have for the gauge fixing factors in (4.6):
\begin{equation}
\tilde{\cal F}_{p} = E^{1 \over 2}_{p+1} E^{-{1 \over 2}}_{p+2} ...=
{\cal F}_{n-p-1}
\end{equation}
and for measure on the space of zero modes:
\begin{eqnarray}
&&d \tilde{\mu}_{0}^{(p+1)} = \prod_{i=1}^{\beta_{p+1}} db_{i}^{(p+1)} det^{1
\over 2}
(<\phi_{i}^{(p+1)}, \phi_{j}^{(p+1}>)\nonumber\\
&&\prod_{k=1}^{n-p-1} \prod_{i=1}^{\beta_{p+k+1}}
df_{i}^{(p+k+1)} [det (<\phi_{i}^{(p+k+1)}, \phi_{j}^{(p+k+1)}>)]^{(-1)^{k}
\over 2}
\end{eqnarray}
where the ghost zero modes $f_{i}^{(p+k+1)}$ are fermions for odd k and bosons
for
even k.
   One gets similar expression for measure of integration over the fermion
field
$\Psi_{p}$:
\begin{equation}
{\cal D} \Psi_{p}= {\cal D} \Psi^{T}_{p} {\cal F}_{p}^{-1} d\mu_{0,f}^{(p)}
\end{equation}
\begin{equation}
{\cal D} \Psi^{T}_{p}= \prod_{n} d \theta_{n}^{(p)} det^{-{1 \over 2}}
(<\alpha_{n}^{(p)}, \alpha_{m}^{(p)}>)
\end{equation}
\begin{eqnarray}
&&d\mu_{0,f}^{(p)} = \prod_{i=1}^{\beta_{p}} d\eta_{i}^{(p)} det^{-{1 \over 2}}
(<\phi_{i}^{(p)}, \phi_{j}^{(p)}>)\nonumber\\
&&\prod_{k=1}^{p} \prod_{i=1}^{\beta_{p-k}}
dg_{i}^{(p-k)} [ det(<\phi_{i}^{(p-k)}, \phi_{j}^{(p-k)}>)]^{(-1)^{k+1} \over
2}
\end{eqnarray}
where the ghost zero modes $g_{i}^{(p-k)}$ are superpartners for ghosts
$f^{(p-k)}_{i}$ and they are bosons for even k and fermions for odd k.
It is easy to see that the expression for ${\cal D} \Psi_{p}$ is similar to
${\cal D} A_{p}$ but all determinant's powers have opposite sign.

   Finally for the integration measure over fermion form $\Psi_{p+1}$ one gets:
\begin{equation}
{\cal D} \Psi_{p+1} = {\cal D} \Psi^{L}_{p+1} \tilde{{ \cal F }}^{-1}_{p}
d\tilde{\mu}^{(p+1)}_{0,f}
\end{equation}
\begin{equation}
{\cal D} \Psi^{L}_{p+1} = \prod_{n} d\tilde{\theta}^{(p+1)}_{n} det^{-{1 \over
2}}
(< \beta^{(p+1)}_{n}, \beta_{m}^{(p+1)}>)
\end{equation}
\begin{eqnarray}
&&{\cal D} \tilde{\mu}^{(p+1)}_{0,f} = \prod_{i=1}^{\beta_{p+1}}
d\eta_{i}^{(p+1)} det^{-{1 \over 2}}
(<\phi_{i}^{(p+1)}, \phi_{j}^{(p+1)}>)\nonumber\\
&&\prod_{k=1}^{n-p-1} \prod_{i=1}^{
\beta_{p+k+1}} dg_{i}^{(p+k+1)} [det(<\phi_{i}^{(p+k+1)}, \phi_{j}^{(p+k+1)}
>)]^{(-1)^{k+1} \over 2}
\end{eqnarray}
where $g_{i}^{(p+k+1)}$ are bosons for odd $k$ and fermions for even $k$.

   Thus one has for quantum average of gauge invariant operator $\cal O$:
\begin{equation}
Z({\cal O}) = \int {\cal D} A_{p} {\cal D} A_{p+1} {\cal D} \Psi_{p}
{\cal D} \Psi_{p+1} {\cal O} \exp{(-S^{(p)})}
\label{4.16}
\end{equation}
Generally all determinants in the integration measure (4.16) depend on
background metric on manifold $M^{n}$ ( through  of corresponding anomalies).
The detail expressions  for metrical variation of determinants one can find
in ~\cite{as78},~\cite{as84}. It is important for us that in total measure
${\cal D} \mu ={\cal D} A_{p} {\cal D} \Psi_{p} {\cal D} A_{p+1}
{\cal D} \Psi_{p+1}$ all determinants are mutually cancelled, so the total
measure ${\cal D} \mu$ is metrical independent.

   In the case when all Laplacians on $M^{n}$ don't have zero modes let us
consider  the partition function:
\begin{equation}
Z= \int {\cal D} \mu \exp{-S^{(p)}}
\label{4.17}
\end{equation}
Inserting $S^{(p)}$ in the form (3.11) and integrating we obtain:
\begin{equation}
Z\equiv R^{(p)} (M) =  \prod_{n} \pm {\lambda_{n}^{(p)} \over
|\lambda_{n}^{(p)}|}
\label{4.18}
\end{equation}
One can check that it is topological invariant. As in  paper ~\cite{ew88}
it is consequence of that the measure ${\cal D} \mu$ is metrical independent
and $Q$-invariant and metrical variation of the action has the form
$\delta_{g} S^{(p)} =\{ Q, X\}$(2.9). Invariant (4.18) is similar to the one
of the Donaldson invariants ~\cite{ew88} and to the Casson invariant
{}~\cite{ew89},
{}~\cite{aj90}. One chooses sign in (4.18) by the same way as in ~\cite{ew88}.
Let us note that it is consequence of (3.8) that we have identity:
$R^{(p)}(M)=R^{(n-p-1)}(M)$.

   Let $U_{p}$ and $U_{n-p-1}$ be two non-intersecting submanifolds of
dimensions $p$ and $n-p-1$ that represent trivial elements in $H_{p}(M)$
and $H_{n-p-1}(M)$ respectively. Then after standard calculations ( see
{}~\cite{gh90}, ~\cite{sw91} ) one obtains that
\begin{eqnarray}
&&<\int_{U_{n-p-1}} \ast dA_{p} \int_{U_{p}} A_{p}> = <\int_{U_{n-p-1}}
\ast \Psi_{p+1} \int_{U_{p}} \Psi_{p}>\nonumber\\
&&= L(U_{n-p-1}, U_{p}) R^{(p)}(M)
\label{4.19}
\end{eqnarray}
where $L(U,V)$ is linking number of $U$ and $V$ ~\cite{gh90}, ~\cite{sw91}.

   Hence not only metric independent action (1.1) gives us linking numbers
under quantization but the action (2.1) which is metrical dependent and
describes topological theory of Witten type leads to the same result.

   In general case when circles $\gamma_{i}^{(k)}, i=1, ..., \beta_{k}$
form basis of k-dimensional homology $H_{k}(M)$ the partition function $Z$
(4.17) is equal to zero due to zero fermion modes. Then only operators with
correspoding fermion number will give nontrivial average ~\cite{ew88}.
One can consider the measure in functional integral (4.17) as measure on
the direct sum of cohomologies of manifold $M$ : $H=\oplus_{k=0}^{n} H^{k}(M)$
{}~\cite{as84}. Then nontrivial quantum average of an operator is result of
integrating over $H$.

   Let us consider the following gauge invariant operator:
\begin{eqnarray}
&&{\cal O} = F[N_{i_{m}}^{m}] \prod_{i=1}^{\beta_{p}} \int_{\gamma_{i}^{(p)}}
\Psi_{p} \prod_{i=1}^{\beta_{p+1}} \int_{\gamma_{i}^{n-(p+1)}}
\ast \Psi_{p+1} \prod_{i=1}^{\beta_{p+1+2k}} \int_{\gamma_{i}^{n-(p+1+2k)}}
\ast g_{(p+1+2k)}\nonumber\\
&&\prod_{i=1}^{\beta_{p-2k}} \int_{\gamma_{i}^{p-2k}}
g_{(p-2k)} \prod_{i=1}^{\beta_{p-2k-1}} \int_{\gamma_{i}^{(p-2k-1)}}
f_{(p-2k-1)} \prod_{i=1}^{\beta_{p+2k+2}} \int_{\gamma_{i}^{n-(p+2+2k)}}
\ast f_{(p+2 + 2k)}
\label{4.21}
\end{eqnarray}
where
\begin{eqnarray}
&&N^{p}_{i_{p}} =\int_{\gamma_{i_{p}}} A_{p}, N_{i_{p+1}}^{p+1} =\int_{
\gamma_{i_{n-p-1}}^{n-p-1}} \ast A_{p+1} \nonumber\\
&&N^{p-2k}_{i} =\int_{\gamma_{i}^{p-2k}} f_{p-2k}, i=1, ...,\beta_{p-2k}
\nonumber\\
&&N_{i}^{p+1+2k}=\int_{\gamma_{i}^{n-p-2k-1}} \ast f_{p+2k+1}, i= 1,...,
\beta_{
p+2k+1}\nonumber\\
&&N_{i}^{p-2k-1} = \int_{\gamma_{i}^{p-2k-1}} g_{p-2k-1}, i=1,...,
\beta_{p-2k-1}
\nonumber\\
&&N_{i}^{p+2k+2} = \int_{\gamma_{i}^{(n-p-2k-2)}} \ast g_{(p+2k+2)}, i=1,...,
\beta_{(n-p-2k-2)}
\label{4.22}
\end{eqnarray}

Function $F[N^{m}_{i_{m}}]$ in (4.20) is an arbitrary function of the form

\begin{equation}
F[N^{m}_{i_{m}}] = \int_{-\infty}^{+\infty}\prod_{m} dP^{m}_{i_{m}}
\tilde{F}[P^{m}_{i_{m}}] \exp{\imath
(\sum_{m} P^{m}_{i_{m}} N^{m}_{i_{m}})}
\label{4.22}
\end{equation}
which one can consider as an integrable function on the space of cohomologies
$H$ (variables $N^{m}_{i_{m}}$ are coordinates in $H^{m}(M)$).

   We introduce the ghost k-forms $f_{(k)}$ and $g_{(k)}$ in (4.20) which are
fermions or bosons in dependence on $k$ and  correspond to the determinants
$E_{k}^{\pm {1 \over 2}}$ in the measure ${\cal D} \mu$.

   In the cohomology space $H^{k}(M)$ one can choose basis $\phi^{(k)}_{i}$:
\begin{equation}
\int\limits_{\gamma_{j}^{(k)}}^{} \phi_{i}^{(k)} =\delta_{ij}, \int\limits_{
\gamma_{j}^{(n-k)}}^{} \ast \phi_{i}^{(k)} =\delta_{ij}
\end{equation}
One can show (by means of direct calculations) that  quantum average of
operator $\cal O$ {(4.20) is metric independent and equal to
\begin{equation}
<{\cal O}>= R^{(p)}(M) \tilde{F}[0]
\end{equation}
where $\tilde{F}[0]$ are Fourier coeffecients  (4.22) at zero. It can be
represented as result of integrating of the function (4.22) over
$H=\oplus_{k=0}^{n} H^{k}$:
\begin{equation}
\tilde{F}[0]= \int\limits_{H}^{}F[N_{i_{m}}^{m}] \prod_{m}^{}dN^{m}_{i_{m}}
\end{equation}

   In the topological theories of Witten  type the quantum averages of
operators after integrating out the non-zero modes  are reprezented as
differential forms over zero modes space. In the Witten model ~\cite{ew88}
it is instanton moduli space. In the model under consideration it is
cohomology space $H=\oplus^{n}_{k=0} H^{k}(M)$.

\section{Non-abelian generalization}
\setcounter{equation}0
   Let $E$ be an vector bundle with base $M$ and group $G$; $\Omega^{p}
(M,E)$ is  space of p-forms taking values in fibres of $E$ and $C$ is
one-form connection in the bundle $E$. Then one has for covariant exterior
derivative of an element $A_{p}\in \Omega^{p}(M,E)$: $d_{C}A_{p} =
dA_{p} + C\wedge A_{p}$. The corresponding generalization of action
(2.1) is following:
\begin{equation}
S^{(p)}= \int\limits_{}^{}{1 \over 2}\ast d_{C} A_{p} \wedge d_{C}A_{p}
+ \ast \Psi_{p+1} \wedge d_{C} \Psi_{p}
\end{equation}
which is invariant under supersymmetry:
\begin{equation}
\hat{Q} A_{p} =\Psi_{p}, \  \ \hat{Q} \Psi_{p+1} =-d_{C}A_{p}
\end{equation}
If $E$ is flat bundle, i.e. $d^{2}_{C}=0$, then the action (5.1) has gauge
symmetry which is similar to (2.2). It is easy to extend analysis of
Sections 2-4 for this case. The essential difference of non-abelian case
 appears in construction of gauge invariant operators. One should insert
the path-ordered exponent $G(x,y)=Pexp\int\limits_{x}^{y} C$. For example
the quantum averages of following operator (which is generalization of
(4.19)):
\begin{equation}
\int\limits_{U_{n-p-1}}^{} \ast dA_{p}(x) G(x,y) \int\limits_{U_{p}}^{}
A_{p}(y)
\end{equation}
will give us the non-abelian generalization of linking numbers ~\cite{gh90}.

   The action (5.1) isn't invariant under  gauge symmetry of type (2.2) if
$E$ is not flat bundle. However we should note  following important
circumstance. The action $S^{(p)}$ (5.1) depends on the external fields:
metric $g$ and gauge field $C$. It is easy to see that variation of $S^{(p)}$
under the gauge field $C$ takes the form of $Q$-commutator:
\begin{equation}
\delta_{C} S^{(p)} = \hat{Q}[ \int\limits_{M}^{}-\ast(\delta C \wedge A_{p})
\wedge \Psi_{p+1}]
\end{equation}
Hence the topological invariants which appear under quantization of (5.1)
will be both metrical invariants (i.e. they are not changed under local
variation of metric $g$) and invariants of bundle $E$ (i.e. they are not
changed under local variation of connection $C$ in the bundle $E$). The
corresponding non-abelian analog of invariant $R^{(p)}(M)$ (4.18)  is one
of this type.

$Acknowledgements$. I would like to thank Yu.N.Obukhov for very useful
discussions. I also thank D.Fursaev for kind hospitality at Joint Institute
of Nuclear Research in Dubna.

\end{document}